# Simulations for the Development of Quantum Computational Devices for HEP

James B. Kowalkowski[1] and Adam L. Lyon[2]
Fermilab Quantum Institute, Fermi National Accelerator Laboratory, Batavia, IL, USA
March 15, 2022

In this white paper,[3] we describe characteristics of tools for classical simulations of quantum computational devices appropriate for High Energy Physics applications.

Quantum simulation capabilities that will be available on quantum computing devices are useful throughout many areas of physics, including digital and analog simulations for quantum field theories, neutrino nucleus scatter studies, spin systems, improved QCD background predictions for collider experiments, non-abelian gauge symmetry, and even optimization, classification and combinatorics problems that appear in analysis. These devices promise to make these studies possible on reasonable timescales even for problems that scale exponentially. Eventually, these devices will allow direct observation of quantum processes within a controlled laboratory environment, or direct coupling of quantum sensors to computing for dark matter searches.

New quantum computing devices that may be suitable for executing HEP-relevant algorithms are emerging. One example is Fermilab's system using superconducting radio-frequency cavities.[4] This multilevel "qudit" device will exhibit extremely long coherence times and could have a major impact on the ability to run complex algorithms. A co-design process for the design and construction of the devices as well as the design of the algorithms will ensure that HEP needs are met. An important component of this process is a simulation system run on classical computers offering fast turn-around time. Such a system allows for quick testing of ideas and strategies as well as understanding the impact of device and algorithm constraints. This system will also provide a good entry point for scientists and students that want to engage in quantum projects without the complexity of interfacing with commercial or proprietary systems.

---

[1] jbk at fnal.gov
[2] lyon at fnal.gov
[3] Contribution to Snowmass 2021
[4] Superconducting Quantum Materials and Systems Center, https://sqms.fnal.gov

The type of simulation codes that are applicable to the problem are categorized as "Open Quantum System Simulations". Open indicates that the device's interactions with its environment and resulting noise/decoherence are taken into account (as opposed to a Closed system where the device is isolated from the environment). With such simulations, it is possible to predict behavior of the device and system under real-world conditions. Open systems simulators will typically evolve the density matrix using a master equation in Lindblad form.[5]

Beyond predicting real-world behavior of a quantum device, applications of such simulations include gate design and pulse building with optimal control. These are especially important for developing algorithms on multi-mode multi-level devices.

Desired functionality of an open quantum system simulator at the scale required for co-design are

- Supports sufficient size parameters - qubits counts, qudits levels, and modes.
- Performant - quick turnaround time, taking advantage of high performance libraries and features including GPU and TPU accelerators, large distributed memory spaces, and high-speed interconnects.
- Friendly interface - Need to use the system as you would a real device from popular programming environments quickly and efficiently using languages such as C++, Python, and Julia.
- Interfaces well with community standard circuit-building tools - Accepts output from standard compilers and standard instruction formats such as QASM[6].
- Can be easily driven from pulse descriptions and sequences generated from standard community tools
- Useful not only for execution of digital algorithms, but also of more direct quantum simulations or experiments that might be carried out.
- Library of device Hamiltonians and device profiles to quickly get started
- Usable from cloud services and HEP batch systems.

There are several available open quantum system tools and toolkits. QuTip[7] is a Python based system. It offers an easy-to-use interface and compatibility with user-friendly tools such as Jupyter Notebooks. However, the downside of its Python roots are lack of performance for simulations of the large scale that are important for device studies. There are several options that offer more performance: C based QuaC[8] and C++ based Quandary.[9] The Julia based

---

[5] See D.A. Lidar, https://arxiv.org/abs/1902.00967
[6] A. Cross et. al., "Open Quantum Assembly Language", https://arxiv.org/abs/1707.03429
[7] J. R. Johansson, P. D. Nation, and F. Nori: "QuTiP 2: A Python framework for the dynamics of open quantum systems.", Comp. Phys. Comm. 184, 1234 (2013) (http://qutip.org)
[8] https://github.com/0tt3r/QuaC
[9] S. Guenther, N.A. Petersson, J.L. DuBois, AVS Quantum Science, vol. 3, https://arxiv.org/abs/2106.09148 (2021) and IEEE Supercomputing 2021 (2021) https://arxiv.org/abs/2110.10310

JuqBox[10] package, though a closed quantum simulation, is useful for designing pulses. The authors of this whitepaper have experience with QuaC, having used it to study the impacts of noise and structure on quantum information encoded in a quantum memory.[11] Many of these codes are capable of running on High Performance Computers or clusters outfitted with accelerators, and may take advantage of GPUs through community supported tools such as PETSc[12].

A further important application of an open quantum system simulator is as part of a quantum device's control system. In such a system, the simulation may be used to understand the impact of noise within the device and electronics as well as testing the fidelity of gates resulting from their constructed pulse sequences.

Most quantum offerings and even some simulators allow for submission of jobs or experiments via the Cloud. Standardizing this submission with a common interface and a complete set of features (e.g. handling data file transfers, queueing, setting up a common environment) is the goal of the Fermilab HEPCloud project.[13] HEPCloud is a system for the submission and management of jobs for processing High Energy Physics experimental data. It allows the user to interact transparently with local computing, Grid computing, Cloud computing, and High Performance Computing resources. It is natural to extend this system to Quantum Computing resources and provide an easy way for the QIS and HEP communities to run experiments on quantum computational devices via the Cloud.  Work has begun to integrate HEPCloud with the Rigetti computing platform.

Controls hardware for quantum computational devices is another area where simulation can play an important role. For example, Fermilab leads the QICK (Quantum Instrumentation Control Kit) project,[14] developing a Xilinx RFSoC based qubit controller. An open systems simulator on the backend of this controller could stand-in for a quantum device, making development easier and less costly. A tandem project aims to simulate the scheduling of pulses produced from QICK firmware and sent to RF drivers. This simulation will allow for validating and debugging output pulse sequences using a mock system instead of tying up actual hardware.

In summary, simulations play a critical role in the development of quantum computational devices as well as their control systems. Tools meeting the requirements specified in this document would be especially useful, allowing for development, testing, validating, and debugging without tying up the in-high-demand hardware. Development activities include design of algorithms and system building blocks such as multi-level gates.  There are several open source packages and toolkits in a variety of computer languages that may be suitable.

---

[10] https://github.com/LLNL/Juqbox.jl
[11] M. Otten, *et. al.,* Phys. Rev. A **104**, 012605 (2021)  https://arxiv.org/abs/2011.13143
[12] https://petsc.org/release/overview/
[13] https://hepcloud.fnal.gov
[14] L. Stefanazzi, *et. al.*, https://arxiv.org/abs/2110.00557

Simulation of the control system electronics may also be fruitful. Since quantum devices are in high demand and costly, simulations serve as a good mock system to research ideas and debug implementations. Their role in Quantum Computing for HEP applications will only increase in the future.